# Quantum calibrated magnetic force microscopy

*Baha Sakar[1,2,‡], Yan Liu[3,4,‡,*], Sibylle Sievers[1], Volker Neu[5], Johannes Lang[3], Christian Osterkamp[3], Matthew L. Markham[6], Osman Öztürk[2], Fedor Jelezko[3], Hans W. Schumacher[1]*

[‡] *These authors contributed equally to the work*
[*] *Corresponding Author*

1. Physikalisch-Technische Bundesanstalt, 38116 Braunschweig, Germany
2. Gebze Technical University, Department of Physics, 41400, Kocaeli, Turkey
3. Institute for Quantum Optics, Ulm University, 89081 Ulm, Germany
4. Beijing Academy of Quantum Information Sciences, Beijing 100193, China
5. Leibniz IFW Dresden, 01069 Dresden, Germany
6. Element Six Global Innovation Centre, Oxfordshire, OX11 0QR, United Kingdom

**Abstract**

We report the quantum calibration of a magnetic force microscope (MFM) by measuring the two-dimensional magnetic stray field distribution of the MFM tip using a single nitrogen vacancy (NV) center in diamond. From the measured stray field distribution and the mechanical properties of the cantilever a calibration function is derived allowing to convert MFM images to quantum calibrated stray field maps. This novel approach overcomes limitations of prior MFM calibration schemes and allows quantum calibrated nanoscale stray field measurements in a field range inaccessible to scanning NV magnetometry. Quantum calibrated measurements of a stray field reference sample allow its use as a transfer standard, opening the road towards fast and easily accessible quantum traceable calibrations of virtually any MFM.

Quantitative nanoscale stray field measurements are a prerequisite for reliable and sound nanomagnetic research [1-3]. However magnetic force microscopy (MFM), the most versatile tool for nanomagnetic imaging [4,5], generally provides qualitative stray field information, only [6]. Common approaches to quantitative MFM (qMFM) [7-10] rely on simplifying assumptions on the magnetic tip [11-13] or the stray field distribution of a calibration sample [9,14-19] which are hard to validate independently [11,10,20]. Few groups made attempts to directly characterize the magnetic stray field distribution of MFM tips through the Hall effect [21,22] and by Lorentz tomography or holography [20,23-24]. Hall sensors, however, have a limited resolution due to the typical Hall cross dimensions of some 100 nm and Lorentz microscopy-based techniques require complex data postprocessing.

Scanning magnetometry with single diamond nitrogen vacancy (NV) centers opens a new path towards quantum-based quantitative nanoscale field measurements [2,25-26]. However, it is comparably slow and doesn't allow quantitative measurements in the relevant field range of many nanomagnetic materials of a few ten to a few hundred mT. Here, we report quantum calibration of MFM (QuMFM) by measuring the MFM tip's stray field distribution using an NV center as a quantum sensor [27]. Using QuMFM we quantitatively measure the stray field distribution of a typical MFM calibration sample with field amplitudes up to 100 mT, opening the road towards quantum traceable calibration of virtually any MFM.

NV measurements of the MFM tip's stray field are performed in a confocal microscope with bottom optical access and an integrated MFM on top (**Figure 1a**). The NV centers were prepared by 2.5 keV $^{15}$N implantation into a thin layer of ultrapure isotopically enriched diamond (99.99 % $^{12}$C) that was epitaxially grown on the (001) surface of electronic grade type IIa bulk diamond (Element Six, Nitrogen concentration < 6 ppb, often <1 ppb, typical NV concentration < 0.3 ppb). For efficient photoluminescence detection the bulk diamond is shaped as a half-sphere structure, forming a solid immersion lens (SIL). Its upper planar surface is oriented along the x-y plane of the MFM. The selected single NV center with a photon rate of 1.2 Mcs$^{-1}$ is situated 3.6 nm below the surface as derived from proton magnetic resonance [28] (s. supplement). The NV axis is oriented along [111] and has a nominal angle $\theta = (54.75 \pm 1)°$ to the surface normal (**Figure 1b**), with the uncertainty resulting from the miscut angle. The MFM tip was first scanned over the NV center in intermittent contact mode over a $2 \times 2$ µm$^2$ area to capture the surface topography. Then, the tip was scanned at a measurement height (i.e. the distance between the tip apex and the surface) of $z_{tot} = 80$ nm above the surface with 100 nm step size and an optically detected magnetic resonance (ODMR) spectrum [29] was measured at each pixel. ODMR measures the shift of the Zeeman-split NV center spin resonance frequencies $f_\pm$ induced by the tip stray field $B^{tip}$ (**Figure. 1c**). The NV spin states are optically accessed by the difference in fluorescence of $m_s=0$ and $m_s=\pm 1$ states. Optical excitation polarizes the $m_s=0$ state and a resonant microwave transfers the population to the „dark" ($m_s=\pm 1$) states with reduced fluorescence. We used π-pulse ODMR for the field sensing, to prepare near 100% spin-population resulting in 30% contrast. The magnetic field sensitivity of the NV center of $0.54 \frac{\mu T}{\sqrt{Hz}}$ is determined by the linewidth of its ODMR spectrum and by the photon collection efficiency. From the resonance frequencies the magnitude of the magnetic induction $B^{tip} = |\vec{B}^{tip}|$ and its angle $\beta = \sphericalangle(\overrightarrow{NV}, \vec{B}^{tip})$ (cp. **Figure 2c**) with respect to the NV axis are derived (see supplement). $B = \mu_0 H$ is given in units T with µ$_0$ the magnetic field constant and $H$ the magnetic field in units Am$^{-1}$. Since in the literature $B$ and $H$ are often both referred to as ´magnetic field´ we will in the following use this term for simplicity.

The quantum calibrated low moment MFM tip (MFM_LM, TipsNano) is coated by 20 nm CoCr with nominal tip radius of 30 nm. Scanning the metallic MFM tip near the NV center leads to a reduced fluorescence allowing to align the tip over the NV center with 200 nm uncertainty. In the intermittent contact mode, a free rms oscillation amplitude of 10 nm was set with a setpoint of 3 nm during the surface scan. For the NV tip field measurements at $z_{tot} = 80$ nm the free oscillation amplitude was reduced to 3 nm.

**Figure 2** shows the stray field distribution of the MFM tip measured at room temperature and zero applied field. $B^{tip}$ (**Figure 2a**) shows a rotation-symmetric maximum with the tip near the NV center sharply dropping towards the edges. In the 2D plot of the field angle β (**Figure 2b**) the white dotted line marks the projection of the NV axis with azimuthal angle $\varphi = 119°$ (**Figure 1b**). β is mirror symmetric around $\varphi$ as expected for a rotation-symmetric tip stray field. The variation of β along the axis is illustrated in **Figure 2e**. It sketches the stray field lines of a point-like tip in the plane defined by the surface normal and the NV axis. The red arrows indicate the NV axis, tilted by $\theta = 54.75°$ with respect to the surface normal, for three different relative positions of tip and NV center. At (1) the field and the NV axis are parallel. The corresponding region (1) on the top left of **2b** is characterized by low angles (dark blue). At (2) the tip is positioned above the NV center with $\vec{B}^{tip}$ perpendicular to the surface and hence β ≙ θ. (3) reveals values around 90° with the field lines almost perpendicular to the NV axis (dark red in **2b**). The observed decay of β beyond (3), where $B^{tip}$ is low is attributed to spurious background fields caused by field contributions from upper parts of the tip or magnetic components in the MFM system. **Figure 2c** shows a map of the field component parallel to the NV axis $B_{\parallel}^{tip} = B^{tip} \cos \beta$. Again, the NV axis and the three configurations of **2e** are marked. At (1), since $\cos β ≙ 1$, $B_{\parallel}^{tip}$ shows large values. Between (1) and (2) a maximum is found. Its position is determined by the competition between the decrease of $\cos \cos \beta$ and the increase of $B^{tip}$ with decreasing distance of tip and NV center. When following the dashed line to the lower right from (2) to (3) $B^{tip}$ decreases strongly with distance leading to a decrease of $B_{\parallel}^{tip}$. In free space the knowledge of the $B_{\parallel}^{tip}(x, y)$ distribution in a 2D plane suffices to calculate all vector components of $\vec{B}^{tip}(x, y)$ (see supplement). **Figure 2d** shows the calculated out-of-plane component $B_z^{tip}$ of the tip stray field which is used to calibrate the MFM measurements [30]. $B_z^{tip}$ has a maximum with the tip situated above the NV center and shows the expected rotational symmetry. Additionally, $B^{tip}$ was measured as function of the tip-surface distance with the tip positioned near the maximum of $B^{tip}$ (**Figure 2f**). The data were measured without cantilever oscillation with the zero tip-sample distance defined by a repulsive tip-sample force of 2.66 nN. The measured field decrease with increasing distance is well described by an exponential decay (dashed line). By extrapolating to zero distance, we derive a maximum tip stray field of around 44 mT at the tip apex. This compares well with stray field estimations from qMFM calibration procedures for similar low moment tips [15].

In qMFM, the measured MFM signal is related to the quantitative stray field distribution by the so-called instrument calibration function, $ICF$, being the calibrated point spread function of the imaging process. A major contribution to the $ICF$ is the tip stray field distribution $B_z^{tip}(x,y)$, which determines the spatial broadening associated with the MFM measurement. In 2D Fourier space the formula governing these relations reads as follows (see supplement [7]).

$$\Delta\Phi(\mathbf{k},z) = ICF(\mathbf{k}) \cdot B_z^{sample}, \text{ with}$$

$$ICF(\mathbf{k}) = \frac{2\,Q}{C \cdot \mu_0} \cdot [LCF(k,\Theta,A)]^2 \cdot k \cdot e^{kz} \cdot B_z^{tip*}(k,0) \tag{1}$$

$Q$ and $C$ are the quality factor and the stiffness of the oscillating cantilever, the $LCF$ corrects for the finite oscillation amplitude $A$ and canting angle $\Theta$ of the cantilever.

The NV data of **Figure 2d** for the first time directly delivers a quantum calibrated $B_z^{tip,NV}(x,y)$. **Figure 3a,b** show linear sections of $B_z^{tip,NV}(x,y)$ along $x$ and $y$ through the maximum of **Figure 2d**. The blue measured data points are shown together with fits (red lines) to the data. The fits are used to extrapolate the measured stray field data to field values outside the 2 μm x 2 μm measurement window to consider longer range magnetic interactions. The according 2D extrapolated stray field distribution (**Figure 3c**) shows a sharp maximum of around 14 mT and a slight distortion of the rotational symmetry mostly resulting from the tilt of the MFM cantilever (**Figure 1a,d**). This NV measured stray field distribution of the tip $B_z^{tip,NV}(x,y)$ in combination with the cantilever's mechanical properties yield the quantum calibrated $ICF$. Using this quantum calibrated $ICF$, any consecutive MFM measurements can be directly converted into quantum calibrated QuMFM stray field maps without requiring knowledge on the sample.

The principle of MFM measurements is schematically sketched in **Figure 1d**. As a test sample we use a Co/Pt multilayer with perpendicular magnetic anisotropy (PMA) that shows perpendicular stripe domains with 170 nm average domain width (for details see supplement). It exhibits a maximum stray field of about 100 mT at a few tens of nm surface distance, a field range typical for many industrially relevant thin films but inaccessible for direct quantitative mapping by scanning NV magnetometry. Furthermore, the sample is a typical reference material for classical qMFM calibration [31]. This sample was chosen since, for this particular PMA material, the sample stray fields can also independently be estimated, based on the well characterized global sample properties, from the MFM phase shift measurement (see Supplement section G and below) and thus it allows to compare its QuMFM stray field maps to an independent source of knowledge. For MFM measurements, the surface topography is first determined by a scan in intermittent contact mode. Then, in a second scan at height $z_{tot}$, the phase shift $\Delta\Phi$ of the cantilever oscillation is sampled. It is caused by the magnetic tip-sample interactions and constitutes the MFM raw data of **Figure 4a**. MFM data were collected in a scan range of (5.11 µm)$^2$ with 512 x 512 pixels. The cantilever spring constant $C = (3.159 \pm 0.442)$ N/m was determined from thermal fluctuations and the resonance quality factor $Q = 160.7 \pm 5$ from the resonant oscillation peak at $z_{tot}$. Two consecutive cycles of NV-calibrations and MFM measurements delivered comparable results speaking for a good stability of the MFM tip under the given calibration and measurement conditions.

Exploiting the relation from eq.1, the QuMFM measured quantum calibrated stray field distribution $B_z^{QuMFM}$ at height $z_{tot} = 80$ nm (**Figure 4b**) is then derived by a deconvolution of the raw data of **Figure 4a** with the quantum calibrated $ICF$ using a pseudo-Wiener filter for regularization. The regularization parameter is chosen such that the instrument noise is filtered without significantly cutting contributions from the stray field spectrum. We would like to note again, that the calculation of the quantitative stray field distribution $B_z^{QuMFM}$ from the measured phase shift data $\Delta\Phi$ merely requires the NV calibrated tip stray field data $B_z^{tip,NV}$, and that no prior knowledge on the sample magnetic properties is needed. Such a calculation is possible for any sample, independent on its magnetization distribution.

**Figure 4g** compares sections of the reconvolved Wiener filtered MFM data (red) with the initial experimental MFM data (green) as a check of consistency. They agree well, speaking for a moderate accuracy loss due to filtering. More details on the data analysis and deconvolution procedures are given in the supplement, section D.

**Figure 4c** shows the estimated reference stray field $B_z^{ref}$ at $z_{tot} = 80$. For the calculation, the MFM raw data is first deconvolved with $B_z^{tip,NV}(x,y)$ to reduce the feature broadening induced by the imaging process and then discriminated, resulting in a well-founded guess of the domain pattern. In a second step, from the associated surface charge pattern (both at the upper and lower surface of the film) and the transition at the domain boundaries, the stray field $B_z^{ref}$ is calculated at the desired height above the sample surface. Details are described in the supplement section E.

**Figure 4d** shows the $B_z^{qMFM}$ data as obtained by conventional qMFM [31]. In a conventional calibration $B_z^{tip}(x,y)$ is not known a priori. Hence, the guess of the reference domain pattern is derived by directly discriminating the raw data of **Figure 4a** (i.e. without prior deconvolution with $B_z^{tip}(x,y)$). This simplified guess of the stray field distribution is used to derive the classical ICF and hence to calibrate the data.

**Figure 4e** compares horizontal sections through **b** and **c** at the center of the images. $B_z^{QuMFM}$ is shown in red and $B_z^{ref}$ in blue. The data agree well within the uncertainty bands (shaded regions) thereby affirming the credibility of the quantum calibration. For the given calibration we derive an uncertainty of $B_z^{QuMFM}$ of $u_{tot} = 20\ \%$ for fields around 100 mT (see supplement). $u_{tot}$ is dominated by the rather large uncertainty of the cantilever's spring constant $C$ ($u_C = 14\ \%$) and the resonance quality factor $Q$ ($u_Q = 3\ \%$). Smaller contributions stem from the uncertainty of the tip-sample distance during tip calibration and MFM measurement, as well as from MFM noise and numerical uncertainties (see supplement). With respect to $u_{tot}$ the field uncertainty of the ODMR data can be neglected. The main uncertainty contribution of the model fields $B_z^{ref}$ of 6 % stems from the sample's saturation magnetic moment $m_s$. **Figure 4f** compares $B_z^{QuMFM}$ (red) and $B_z^{qMFM}$ (black). The agreement of the data within the uncertainty bands can be considered the first independent validation of the classical qMFM calibration. However, since the $ICF$ of qMFM is derived from the MFM image without prior knowledge of the broadening of magnetic features by the tip [30], the $ICF$'s sharpness is overestimated and hence narrow stray field features are suppressed. An example illustrating this effect is marked by the ellipses in **Figure 4b-d**, where a narrow domain feature found in $B_z^{QuMFM}$ and $B_z^{ref}$ is not revealed in $B_z^{qMFM}$.

Quantum calibrated QuMFM as demonstrated here has several advantages over qMFM. Besides being quantum traceable, it does not rely on a reference sample and its properties. In qMFM, the $ICF$ calibration is only reliable in the range of spatial wave vectors $k$ of the reference sample's magnetic features. Larger and smaller structures cannot be quantitatively measured. In QuMFM, the NV measurement of the tip's stray field can be performed over high-density spatial grids for sampling large-$k$ data, but also over large areas below the tip apex for quantifying the low-$k$ contributions of $B_z^{tip,NV}$. Thus, QuMFM with a such calibrated tip can be applied to magnetic stray field landscapes containing any length scales, also combining nanometer and micrometer feature sizes. The smoothness of the tip's $B_z^{tip}$ profile even allows using an adapted, non-uniform mesh in the NV measurements, which renders such multiscale calibration feasible. This will in the future enable quantitative studies of magnetic multiscale phenomena, e.g. interaction domains in nanocrystalline permanent magnet samples [42], which cannot be analyzed quantitatively, so far. Additionally, the quantitative knowledge of the tip stray field allows evaluating its backaction on the domain structure of the magnetic sample and will enable quantitative investigations of tip field induced manipulation of nanomagnetic objects [33-34]. Furthermore, this knowledge makes it possible to select a suitable type of tip which will interact with the sample in a negligible way concerning its impact on the sample's magnetization.

Note that in the future significantly improved QuMFM field uncertainties can be foreseen. Especially, the dominating uncertainty of the cantilever spring calibration could be drastically reduced to < 1 % by proper calibration [35-36]. Vacuum MFM with orders of magnitude higher Q factor and thus better signal-to-noise should allow a further reduction of the total uncertainty down to about 1 % or less.

Obviously, the tip calibration procedure demonstrated above is too time consuming to be performed routinely for every day's MFM use. Here, the most promising route to quantum traceable MFM measurements is to provide quantum traceably calibrated transfer standards. Note, that the data of **Figure 4b** also represents the first quantum traceable calibration of the stray field of a Co/Pt qMFM reference sample which, with a domain transition width of 15 – 20 nm and a typical domain size of about 200 nm has been successfully used in various studies [9,15,37] for conventional qMFM on magnetic feature sizes from 30 nm to about 1 µm. In the future, qMFM can be based on the *NV measured* QuMFM stray field distribution $B_z^{QuMFM}$ circumventing the systematic errors of $B_z^{ref}$ discussed above. This will turn reference sample based qMFM calibrations into quantum traceably calibrated stray field measurements. Additionally, the QuMFM calibration will allow developing new reference samples with different magnetic properties tailored to specific applications and field ranges since the reference sample's properties are no longer dictated by the hard requirement of a calculable domain model. This opens the way for widely available quantum traceable nano-scale magnetic measurements over a broad field range using virtually any MFM.

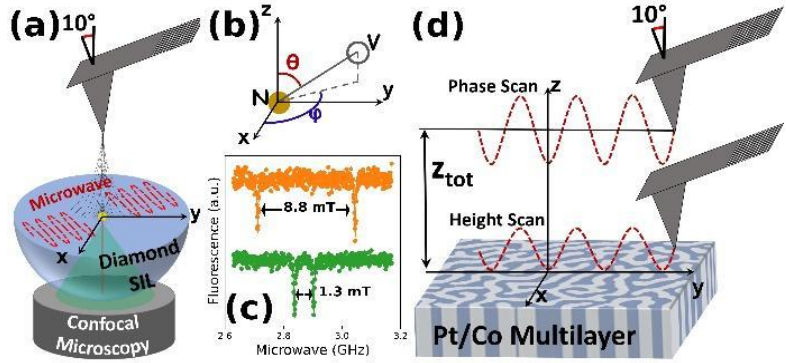

**Figure 1. Experimental Setup**. **a**, NV based stray field measurements of an MFM tip. When scanning the MFM tip over the sample an ODMR spectrum is measured at every point. **b**, Coordinate system of the measurements. The orientation of the NV-axis is given by the polar angle $\theta$ and the azimuthal angle $\varphi$. **c**, Typical ODMR spectrum for two different tip stray fields (fluorescence signal as function of microwave frequency). The curves are offset for clarity. The magnitude of the tip stray field at the position of the NV center and field angle $\beta$ with respect to the NV center axis are derived from the splitting and offset of the resonance lines. **d**, Principle of MFM measurements comprising a topography scan in intermittent contact mode and a phase scan at a tip-sample-distance $z_{tot}$ of 80 nm. The phase signal of the cantilever oscillation as a function of lateral tip position is collected as the MFM raw data.

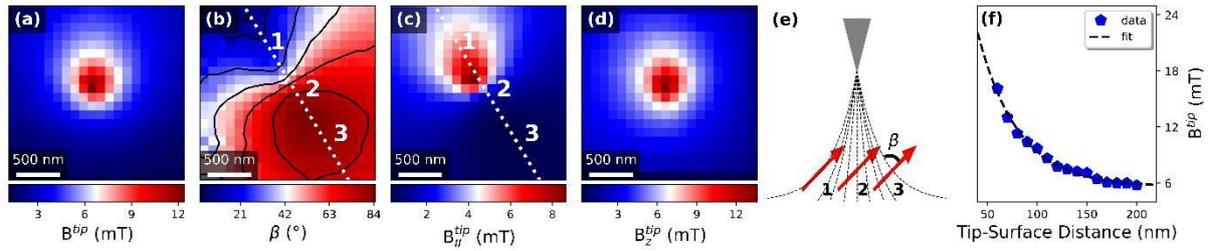

**Figure 2. Characterization of MFM tip stray field by NV magnetometry.** The tip is scanned at $z$=80 nm above the diamond surface. **a** Two-dimensional (2D) map of magnitude of the tip field $B^{tip}$. **b** 2D map of field angle $\beta$ relative to the NV axis. The symmetry axis (dashed line) corresponds to the in-plane orientation of the NV axis. **c** map of $B_{\parallel}^{tip}$. **d** 2D map of the derived z-component $B_z^{tip}$ of the tip stray field. The 2D distribution of $B_z^{tip}(x,y)$ is the key ingredient for quantum calibration of the MFM. **e** Sketch of the relative angle $\beta$ of the tip stray field and the NV axis. **f** Tip-surface distance dependence of $B^{tip}$ with the tip positioned above the NV center. From interpolation (dashed line) a maximum tip stray field of 44 mT at the tip apex is derived

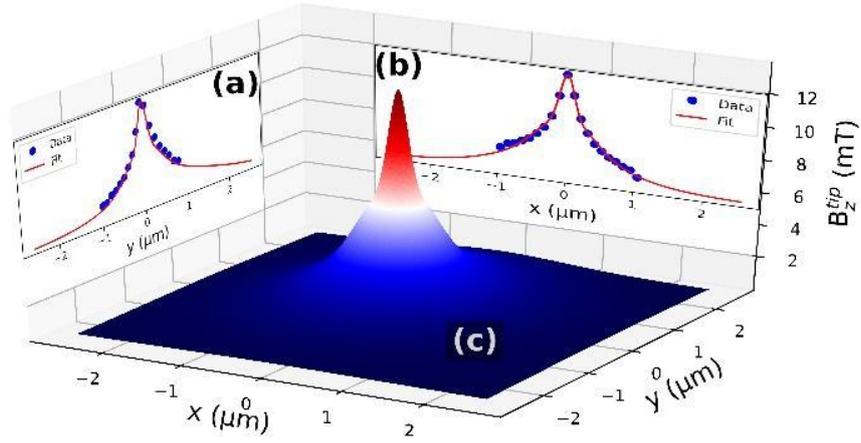

**Figure 3. Quantum calibrated tip stray field distribution. a, b** Sections in x- and y-direction through the maximum of the $B_z^{tip}$ data of Fig. 2e (blue solid symbols). The red lines show combined Gaussian-Voigt fits to the data. **c** Quantum calibrated tip stray field distribution $B_z^{tip,NV}(x,y)$ based on the extrapolated data fits of **a**, **b**. The fits are used to extrapolate the NV measured stray field to an area of 5.11 µm x 5.11 µm to consider longer range magnetic interactions between tip and sample.

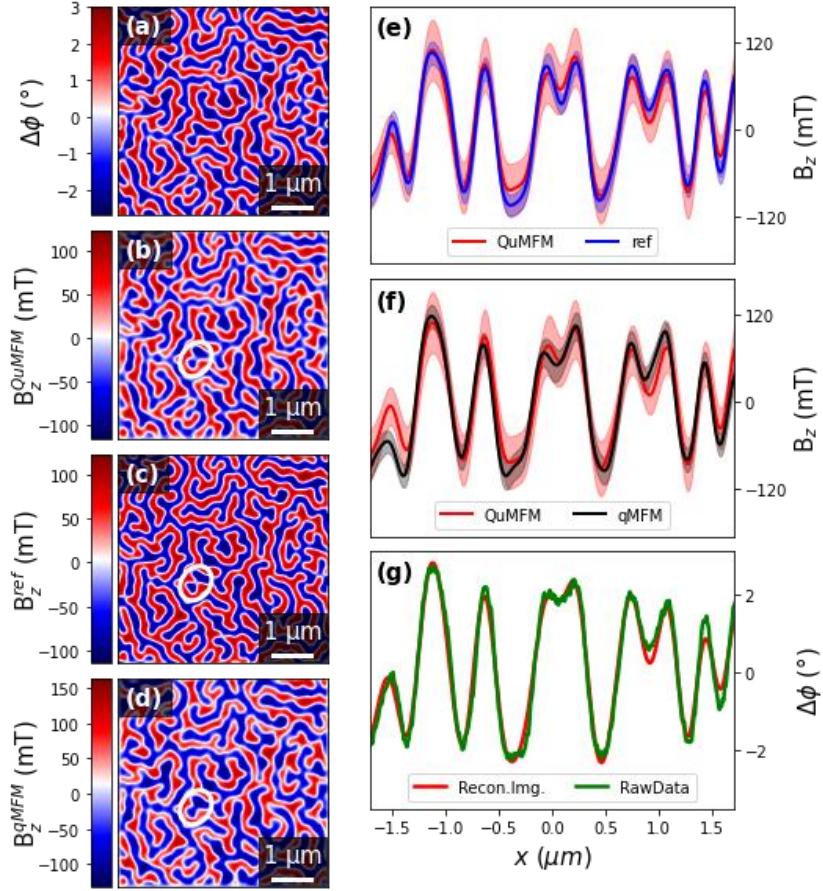

**Figure 4. Quantum calibrated MFM**. **a** MFM data of the Co/Pt multilayer sample with characteristic stripe domains. **b** Quantum calibrated $B_z^{QuMFM}$ at a tip surface distance of 80 nm, obtained from a deconvolution of the MFM data with the instrument calibration function based on the quantum calibrated $B_z^{tip,NV}$ of Fig. 3c. **c** Calculated $B_z^{ref}$ based on the discriminated domain pattern after deconvolution with $B_z^{tip,NV}(x,y)$. **d** Classical qMFM image of the sample's stray field obtained from a deconvolution of the MFM data with a conventionally estimated instrument calibration function. The white ellipses in **b-d** mark characteristic differences of quantum calibrated QuMFM and classical qMFM (see text). **e,f** Comparison of horizontal cross sections of $B_z$ of **b-d** in the center of the images. Uncertainty bands are plotted as shaded regions. **e** Comparison of $B_z^{QuMFM}$ (**b**, red) and $B_z^{ref}$ (**c**, blue). The data agree well within the uncertainty. **f** Comparison of $B_z^{QuMFM}$ (**b**, red) and $B_z^{qMFM}$ (**d**, black). The data again agree within the uncertainty bands. **g** Test of regularized deconvolution procedure applied in **b**. The raw MFM data and the reconvolved data show very good agreement.

**Acknowledgements**

The work was co-funded by the Deutsche Forschungsgemeinschaft under Germany's Excellence Strategy – EXC-2123 QuantumFrontiers – 390837967. The work was supported by European Research Council, European Union's Horizon 2020 research project ASTERIQS, Baden-Württemberg Stiftung and Bundesministerium für Bildung und Forschung.

B.S. acknowledges support from TÜBİTAK-2214-A - International Research Fellowship Programme for PhD Students-1059B141800226 . We thank C. Krien for preparing the Co/Pt multilayer sample.

**Supplemental Information**

Supplemental information includes the detailed calculations of the NV measurements, deconvolution algorithm used for MFM measurements, magnetic field calculations and conversion of the field components, reference and test sample stray field calculations, mechanical properties of cantilever, uncertainty estimations and repetitive measurement consistencies.

**Author Contributions**

Y.L. and B.S. developed the experimental protocol for the tip-calibration and performed the MFM and NV experiments at Ulm. Y.L. built the AFM-confocal setup and provided the tip-field data. B.S. carried out the numerical data analysis to derive the quantum calibration function and the field distributions including uncertainty analysis. B.S., V.N., and S.S. discussed the numerical analysis and the calibration procedures. V.N. provided and characterized the Co/Pt sample. M.L.M. provided the diamond SIL sample. C.O. carried out diamond layer overgrowth on the planar surface. J.L. carried out NV center implantation. F.J., H.W.S., and O.Ö. initiated and coordinated the project. B.S., Y.L., S.S., V.N., F.J., and H.W.S. jointly discussed all data and wrote the manuscript.

# Supplemental Information

**A) NV measurements of the magnetic field using ODMR**

The resonance frequencies of the $m_s = 0 \leftrightarrow \pm 1$ spin transitions can be derived from the spin Hamiltonian of the NV center

$$H = \mu_B g \vec{B} \cdot \vec{S} + D\left(\frac{S_z^2 - S(S+1)}{3}\right) + E(S_x^2 - S_y^2) \quad (1)$$

where $D$ and $E$ are the zero-field splitting parameters, $S = 1$, $\mu_B$ is the Bohr magneton, $g$ is the g-factor and $B$ is the external magnetic field. It is convenient to describe the external magnetic field by its magnitude $B$ and the two angles, $\theta'$ (polar) and $\varphi'$ (azimuthal) with respect to the NV center axis. $D$ and $E$ can be derived from a measurement of an ODMR spectrum in zero field. Having these parameters at hand, $B$, $\theta'$ and $\varphi'$ can be obtained from analysis of the ODMR spectrum as solutions of the characteristic equation (1):

$$x^3 - \left(\frac{D^2}{3} + E^2 + b^2\right)x - \frac{b^2}{2}(D\cos(2\theta') + 2E\cos(2\varphi')\sin^2\theta') - \frac{D}{6}(4E^2 + b^2) + \frac{2D^3}{27} \quad (2)$$
$$= 0$$

where $b = \mu_B g B$. With definition of the energy of the $S_z = 0$ state in frequency units to be $x_0$, the positions of the $S_z = \pm 1$ states are given by $x_\pm = x_0 + v_{0\pm}$, where $v_{0\pm}$ are the experimentally measured frequencies of $0 \leftrightarrow \pm 1$ spin transitions. In strain-free samples, as the SIL used in this work, $E$ is neglectable and $\varphi'$ cannot be determined from the ODMR spectra.

The high photon rate of the NV⁻ center in this diamond SIL sample benefits a high sensitivity of magnetic field measurements, as the minimum detectable static magnetic field by a NV⁻ is given by:

$$B_{min} \approx \frac{2\pi}{\gamma_e} \cdot \frac{\lambda}{c \cdot \sqrt{N}} \quad (3)$$

where $\gamma_e$ is the electron gyromagnetic ratio, $\lambda$ is the full width at half maximum of the measured ODMR resonance, $c$ is the dip contrast, $N$ is the photon rate. In this project, $\lambda \approx 5.0$ MHz, $c \approx 30\%$, and $N \approx 1.2 \times 10^6$ counts/s, and $B_{min} \approx 0.54 \frac{\mu T}{\sqrt{Hz}}$.

**B) Depth measurement of the NV center**

NV centers were prepared in an isotopically enriched epitaxial diamond layer consisting of 99.99% 12C to prolong the spin coherence time and thus improve the sensitivity of the NV center. The thin layer was grown by chemical vapor deposition in a nominally nitrogen free environment. The low nitrogen concentration of this type of samples has been confirmed by electron paramagnetic resonance (EPR), however without providing information on the depth. However, since diamond samples with and without an overgrown layer yield comparable results we estimate the background nitrogen concentration to be on a comparably low level as in the diamond SIL substrate.

With dynamical decoupling (DD) techniques, the spin coherence of a NV center can be prolonged against ambient noises. In addition, with the DD sequence, the alternating magnetic field with its particular frequency can be detected. The NMR signal which essentially is an alternating magnetic field, can be measured by adjacent NV centers with the XY8 sequence, which is one of the commonly used DD sequences [28, 38-42].

And with its limited spatial sensing volume (sensing radius ~ 10 nm) of NMR signals, and depending on the power spectral density of the external NMR spectrum, the depth of a single NV center can be determined [28,41]. In this study, we adopted the depth measurement method by C. Mueller et al. [28]. The detailed method descriptions can be found in Ref. [28], which is open-source.

We applied a constant magnetic field of 63 mT along the NV axis. The Rabi oscillation was measured, and based on the Rabi period, an appropriate XY8 sequence could be generated to sense the proton NMR signal from the external immersion oil which has abundant Hydrogen nuclei. Corresponding to the known proton gyromagnetic ratio of 42.576 MHz/T, the proton NMR signal showed up at 2.7 MHz in the XY8 measurement. Using the known hydrogen density of $50 \times 10^{27}/m^3$ and the measured NMR field, the depth of this NV center is calculated to be 3.54 nm when superposition state was initialized by $\pi/2$ pulse, and 3.66 nm when superposition state was initialized by $3\pi/2$ pulse, as shown in Fig. 1. Hence the depth of the NV center is determined as 3.60±0.06 nm.

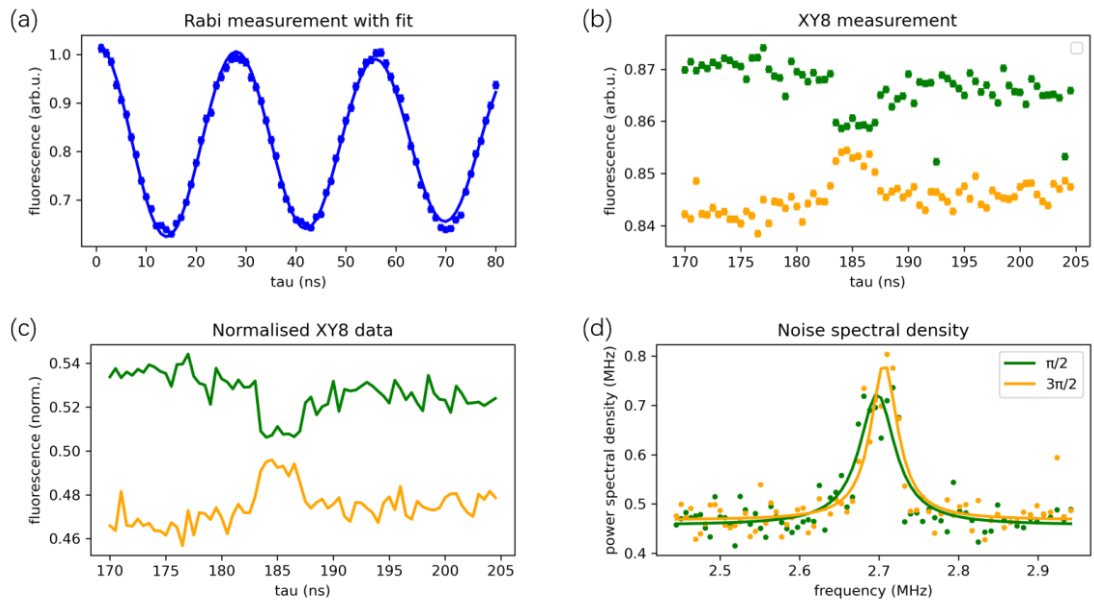

**Figure S1.** Measurement of NV depth by using the external hydrogen NMR spectra. (a) Rabi oscillation measurement; (b) XY8 DD measurements, green dots are measured data where the superposition state was initialized by $\pi/2$ pulse, orange dots are measured data where the superposition state was initialized by $3\pi/2$ pulse; (c) Normalized XY8 data basing on the Rabi oscillation contrast; (d) The inversion of our normalized XY8 data shows the proton NMR signal, and from this spectrum, the depth of NV can be calculated using the equations and parameters discussed in Ref. [28].

## C) Determination of tip-surface distance

With a so-called "force spectroscopy" measurement, which measures the cantilever deflection while approaching the cantilever with a tip towards a hard object surface, the sensitivity of the AFM setup can be measured, as shown in Figure S2. Moreover, what was concerned in our study was how much offset did the "Van Der Walls force" cause to the tip-surface distance.

To ensure well defined contact when approaching the MFM tip onto the diamond surface, the z-piezo was set to slightly over-push the tip towards the diamond surface to reach a repulsion force of 2.66 nN. Converted from this repulsion force, the over-push distance is about 0.7 nm. The 2.66 nN was used as a setpoint value in the study for controlling every tip-surface distance in contact mode. For example, in the stray field measurement, at one designated pixel, the tip firstly approached the diamond surface with this setpoint, and then it was retracted by 80 nm and held still by z-piezo, and then one ODMR spectrum was measured. And then the xy-piezo moved the tip to the next pixel and repeated the same measurement process. And the same procedure was applied to all 400 pixels.

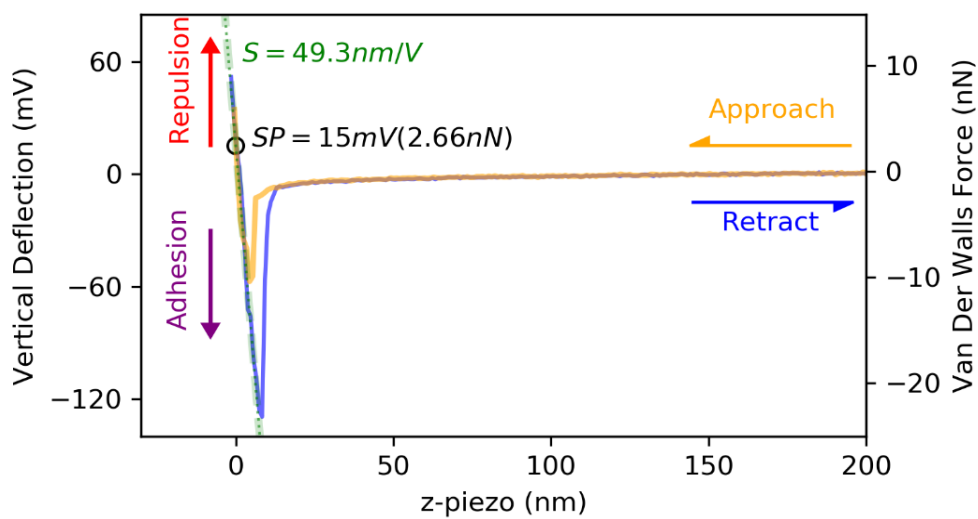

**Figure S2.** Plot of approach (yellow) and retract (blue) curves of the MFM cantilever in contact mode on the diamond sample surface in air. "S" in green indicates the cantilever sensitivity,

and "SP" in black indicates the setpoint value to contact tip apex with diamond surface that was used in the study.

**D) Deconvolution algorithm for MFM measurements**

The image formation in MFM results from an interaction of the magnetic field of the magnetically coated tip, $\overrightarrow{H^{tip}}$, with the magnetization of the sample. The $z$-component of $H_z^{tip}$ together with the mechanical properties of the cantilever determines the instrument's point spread function. The generation of the MFM phase shift signal $\Delta\Phi$ at a measurement height $z'$ as a function of $B_z^{tip} = \mu_0 H_z^{tip}$ at the sample surface and the sample's effective magnetic surface charge $\sigma_{eff}$ distribution can be conveniently described in 2D Fourier space, where the in-plane coordinates $(x, y)$ are mapped to the reciprocal space $(x, y) \rightarrow (k_x, k_x) = \boldsymbol{k}$, with $k = \sqrt{k_x^2 + k_y^2}$, while the $z$-component is retained (partial Fourier space) [43].

$$\Delta\Phi(k, z') = \frac{Q}{C} \cdot \sigma_{eff}^*(k) \cdot [LCF(k, \Theta, A)]^2 \cdot k \cdot B_z^{tip}(k, 0) \qquad (4)$$

The factor $[LCF(k, \Theta, A)]^2$ corrects the impact of the canting of the cantilever and of the finite oscillation amplitude $A$. $Q$ and $C$ are the cantilever oscillation quality factor and stiffness constant, respectively. The asterisk denotes the complex conjugate. Multiplying a function in partial Fourier space with $k$ is equivalent to forming the z-derivative. The factor

$$ICF(k, z') = \frac{Q}{C} [LCF(k, \Theta, A)]^2 \cdot k \cdot B_z^{tip}(k, 0) \qquad (5)$$

establishes the Instrument's point spread function, the so-called Instrument calibration function. $\sigma_{eff}(k)$ can thus be calculated from the measured $\Delta\Phi(k, z')$ image by a deconvolution with the $ICF$. Here, a regularized deconvolution is implemented in the form of a pseudo-Wiener filter [44] with regularization parameter α to avoid the amplification of noise from the measured data:

$$\sigma_{eff}(k) = \frac{1}{k}\frac{C}{Q} \cdot \frac{\Delta\Phi(k,z')}{[LCF(k,\Theta,A)]^2} \cdot \frac{B(k,0)^*}{|B(k,0)|^2 + \alpha} \tag{6}$$

The effective magnetic surface charge $\sigma_{eff}$ distribution describes a virtual magnetic charge density at the sample surface generating the same magnetic stray field in the space above and below the sample as the sample's 3D magnetization distribution. For samples with thickness independent magnetization and no volume charges, surface charges at the lower and upper surface of the sample can be defined by a projection of the sample magnetization $\vec{M} = Ms\,\vec{m}$ onto the surface normal vectors, $\vec{s}^{+/-}$ as $\sigma^{+/-} = \vec{M} \cdot \vec{s}^{+/-}$ (Ms is the saturation magnetisation and $\vec{m}$ the unit magnetization vector). The total effective surface charge pattern $\sigma_{eff}$ at $z' = 0$ is then derived from a projection of the lower surface charges to the upper sample plane and by including a Bloch type domain transition with a width of 16nm in the form of a convolution with a domain wall operator $DW$ [46]:

$$\sigma_{eff}(k) = \sigma^+(k) \cdot Ms(1 - e^{-kd}) \cdot DW(k) \tag{7}$$

### E) Magnetic field calculation from an effective charge density distribution

The sample's magnetic field at a height $z'$ above the surface can be calculated from the sample's $\sigma_{eff}(k)$ in Fourier space using a field transfer function

$$B_z^{sample}(k,d) = \frac{1}{2}\mu_0 \cdot e^{-kz'} \cdot \sigma_{eff}(k) \tag{8}$$

Equation 4 and 8 can be combined to relate measured phase shift data to the sample stray field rather than to the sample's effective magnetization:

$$\Delta\Phi(k,z') = \frac{2Q}{C \cdot \mu_0} \cdot B_z^{sample}(k,z') \cdot e^{kz} \cdot [LCF(k,\Theta,A)]^2 \cdot k \cdot B_z^{tip*}(k,0) \tag{9}$$

### F) Conversion of field components

To calculate $B_z^{tip} = \mu_0 \cdot H_z^{tip}$ from the projection of the magnetic field $B_\parallel^{tip} = \mu_0 \cdot H_\parallel^{tip}$ onto the NV axis, we exploit the fact that in the space outside the tip, where $\nabla \times H = 0$, a scalar potential $\Phi$ exists, with $H = -\nabla \Phi$. Therefore, the complete field vector $\vec{H} = (H_x, H_y, H_z)$ can be calculated from a single component, e.g., from $H_z$. In partial Fourier space $\vec{H}$ is given by $H = (-\frac{ik_x}{k}, -\frac{ik_y}{k}, 1) \cdot H_z$ [43], and consequently, the field component $H_\parallel$ along the NV-axis described by its normalized axial vector $\vec{n}$ is calculated as [30]

$$H_\parallel = \vec{n} \cdot \left(-\frac{ik_x}{k}, -\frac{ik_y}{k}, 1\right) \cdot H_z \tag{10}$$

which can be solved for $H_z^{tip}$ and thus for $B_z^{tip}$.

**G) Reference stray field calculations**

For the well characterized Co/Pt reference samples the magnetic stray field landscape $B_z^{ref}$ is calculated via domain theory from the observed domain configuration. To this end, the measured MFM data are first deconvolved using the calibrated $B_z^{tip,NV}$ to reduce the smearing impact of the extended tip. A threshold criterium (equal maximum signal of up and down band domains in the demagnetized sample state) is then applied to dichotomize the resulting deconvolved image into domains with normalized magnetization $\vec{m} = (0,0,\pm 1)$. The effective surface charge density $\sigma_{eff}^{ref}$ is calculated following Equation (7) using $M_s = m_s/(A \cdot d) = $ (502±30) kA/m as the saturation magnetization where $m_s$ is the sample total magnetic moment, $A$ the sample surface area and $d$ the thickness (132 nm) of the magnetic material.

**H) Conventional qMFM analysis**

The magnetic domain sample used in this study is constructed by Co/Pt multilayer, which is a typical reference material for classical qMFM calibration. It was prepared by magnetron sputtering on thermally oxidized Si(100) with a layer sequence from top to bottom of Pt(2 nm)/[(Co(0.4 nm)/Pt(0.9 nm)]$_{100}$/Pt(5 nm)/Ta(5 nm) resulting in a strong perpendicular magnetic anisotropy of $K_u$ = (0.425 ± 0.025) MJm$^{-3}$. At zero applied field perpendicular stripe domains form with 170 nm average domain width separated by Bloch domain walls with width of $\delta_{DW}$ = 16 nm determined from domain theory. The saturation moment of $m_s$ = (9.3±0.4) 10$^{-7}$ Am$^2$ was determined by vibrating sample magnetometry and the total thickness of (132±3) nm was confirmed by x-ray reflectometry.

To calculate effective surface charges $\sigma_{eff}^{qMFM}$, and eventually stray field patterns $B_z^{qMFM}$ from a conventional qMFM measurement and analysis, the same deconvolution process as in E, Eq. 6 is applied. In qMFM, however, the $ICF$ is derived from a prior calibration of the $ICF$ based on a deconvolution of an additional MFM measurement with the corresponding estimated effective charge density distribution. To this end, the MFM response of a calculable reference sample is measured and the surface charge pattern $\sigma_{eff}^{ref,qMFM}$ is calculated, as discussed in D, however without prior deconvolution step, Eq. 4 can then be applied to derive at the $ICF$ by deconvolving the measured phase shift $\Delta\Phi(k)$ with the reference surface charge $\sigma_{eff}^{ref,qMFM}$. This approach is one of the standard methods in qMFM to calibrate MFM sensors [14, 43] and is justified for well characterized samples such as the Co/Pt multilayer, but it relies on domain theory to connect global magnetic sample properties with local magnetic stray field values and is thus not quantum traceable.

**I) Cantilever mechanical properties**

The calculation of the $ICF$ (equation (5)) requires knowledge on the mechanical properties of the tip cantilever in the form of the tip oscillation quality factor $Q$ and the cantilever stiffness $C$.

The cantilever stiffness $C$ is measured by using the thermal noise spectrum of the cantilever motion on the same SPM as used in the measurements. The free fluctuation of the tip is plotted against the frequency and the curve is fitted by using a simple harmonic oscillation function. The area is used to calculate the total energy of the system in the resonance. By using the equipartition theory, spring constant $c$ is calculated from the total energy.

$$E_{Ther} = \frac{1}{2}k_B T = \frac{1}{2}C\langle q^2 \rangle \tag{10}$$

The quality factor $Q$ is calculated by fitting the resonance curve of the tip recorded near to the surface (~100 nm). $Q$ is determined by calculating the full width of the resonance curve at 0.707 of the maximum amplitude of the resonance peak [31,47].

**J) Uncertainty estimations**

To estimate the Type $B$ uncertainties of the calculated and simulated field distributions, we propagated the uncertainties of the measurement data and measurement parameters through the deconvolution process following a GUM [48] conform approach. The resulting uncertainty describes the expanded combined standard uncertainty with a coverage factor of 2. For the calculation steps that are performed in Fourier space, the uncertainties of the incoming 2D-distributions (tip stray field $B_z^{tip}(x,y)$, $\Delta\Phi(x,y)$, are firstly propagated from real to Fourier space [43, 47]. In the case (i) of the calculation of the quantum calculated $B_z^{QuMFM}$ the uncertainties are then propagated through the regularized deconvolution (equation (6)) and through the multiplication with the wave vector matrix $k$. Furthermore, the uncertainties of the tip's mechanical properties that give the multiplicative factor $Q/c$ are included (equation (4)). Additionally, the uncertainties of the area element in the Fourier transform $dx \cdot dy$, resulting from the SPM positioning uncertainty, are regarded. In the case (ii) of the calculation of the simulated field, $B_z^{cal}$, the uncertainties of the sample magnetic moment per area $m_s$, thickness $d$ and of the measurement height $z_{tot}$ are propagated through the combined equation (7) and equation (8). The used uncertainty data are summarized in **Table S1**.

| Parameter | Symbol | Used value | Uncertainty |
|---|---|---|---|
| 1) calibrated field measurement | | | |
| MFM phase shift data | $\Delta\Phi$ | $\Delta\Phi(x,y)$ | $u(\Delta\Phi) = 0.03°$ [1] |
| Tip stray field | $B_z^{tip}$ | $B_z^{tip}(x,y)$ | $u(B_z^{tip}) = 4\,mT \cdot B_z^{tip}/max(B_z^{tip})$ [2] |
| Measurement height | $z_{tot}$ | 80 nm | $u(z_{tot}) = 10$ nm [3] |
| Cantilever oscillation quality factor | $Q$ | 160.7 | $u(Q) = 5$ [4] |
| Cantilever stiffness | $C$ | 3.159 N/m | $u(C) = 0.442$ N/m [5] |
| Regularization parameter | $\alpha$ | 2.62 x 10$^4$ | $u(\log_{10}(\alpha)) = 1$ [6] |
| 2) simulated sample stray field | | | |
| Sample areal magnetic moment | $m_s/A$ | 66.264E-6 kA | $u(m_s/A) = 3.969$E-6 kA [7] |
| Sample magnetic material thickness | $d$ | 132 nm | $u(d) = 4$ nm [7] |

**Table S1.** Parameters and uncertainties used to calculate the type B uncertainties of the quantum calibrated stray field $u(B_z^{QuMFM})$ and of the calculated and reference stray fields $u(B_z^{cal})$, $u(B_z^{cal})$ of the Co/Pt test sample.

(1) The uncertainty of $\Delta\Phi$ is calculated as the standard deviation of a noise measurement of a non-magnetic sample with the same SPM as used for the calibrated measurement.
(2) The uncertainty of the quantum calibrated tip stray field $B_z^{tip}$ is estimated from comparing it to the measured $B^{tip}$ ($B_z^{tip}$ should everywhere be lower than the maximum of $B^{tip}$). The uncertainty here is relatively high due to numerical features arising from the limited measured area of $B^{tip}$ and due to signal leakage during the involved DFT process. It can be (and will be in future measurement) significantly reduced by choosing an adapted measurement grid for $B^{tip}$.
(3) The uncertainty of $z_{tot}$ is pessimistically estimated from the tip oscillation amplitude calibration process.
(4) The uncertainty of $Q$ is estimated from the width of the fit to the cantilever resonance curve.
(5) The uncertainty of the cantilever stiffness is estimated by using nanoforce measurements. The cantilever stiffness of a similar tip from the same batch and a different commercial brand tip is measured by using thermal noise technique on the SPM then by nanoforce measurement at "PTB Nanonewton Force facility" [13]. The latter technique is destructive so it could not be used for the actual tip used in the study. The relative error between both techniques is found to be 14% which is used as the uncertainty of the cantilever stiffness.
(6) The uncertainty of the regularization parameter is estimated from the discrepancy of the used $\alpha$ and $a^L$ as resulting from an L-curve criterion [49].
(7) The uncertainties of the properties of the measured sample, $m_s A^{-1}$ and $d$ are given by the uncertainties of the respective characterization techniques, VSM and x-ray reflectometry.

## K) Repetitive measurement

The NV and MFM measurements have been carried out two times showing comparable results. Figure S1 shows the data of $B_z^{tip}$ derived from the two experimental runs in (a) and (b), respectively. After each NV measurement an MFM measurement on the Co/Pt sample was performed. The data in (b) corresponds to the data discussed in the main text and shown in **Figure 2e**. The lower panels show for comparison sections through the field maximum of the two NV measurements. The two data sets show a good agreement speaking for a sufficient stability of the MFM tip during the NV and MFM experiments and a relatively low tip wear under the given measurement conditions.

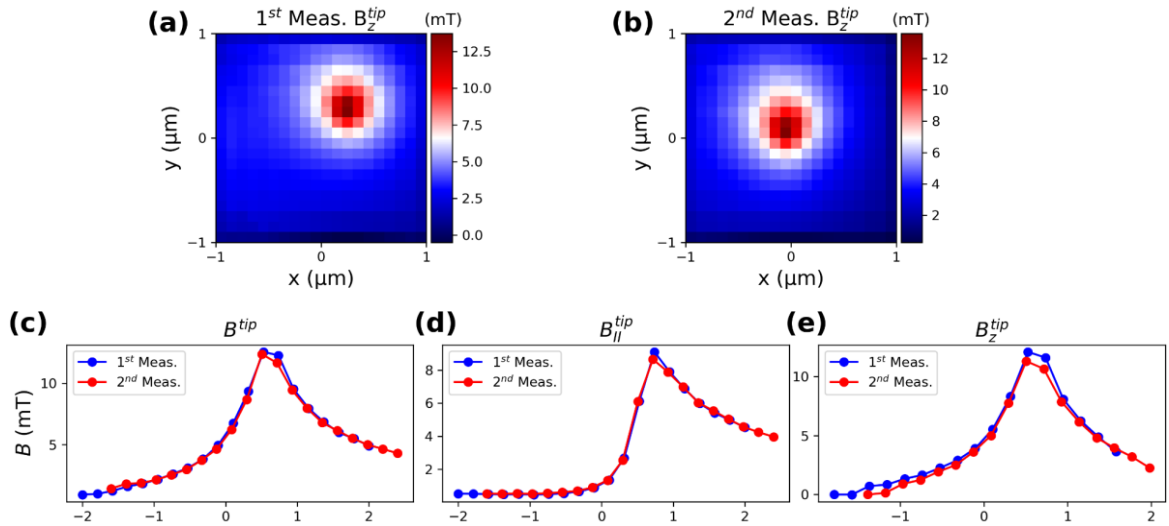

**Figure S1. Repetitive NV measurements. a, b** shows 2D maps of $B_z^{tip}$ for the two experimental runs. **c - e** show comparisons of the field data of the two NV scans of the tip. Line scans in x-direction through the maximum of $B^{tip}$. $B_{||}^{tip}$, $B_z^{tip}$, and $B_z^{tip}$ show a good agreement for the two calibration runs.